# A Tentative Modeling Study of the Effect of Wall Reactions on Oxidation Phenomena

*Richard Porter\*, Pierre-Alexandre Glaude, Frédéric Buda, Frédérique Battin-Leclerc[†]*

Département de Chimie-Physique des Réactions, Nancy Université, CNRS, ENSIC, 1 rue Grandville, BP 20451, 54001 NANCY Cedex, France

Richard.Porter@ensic.inpl-nancy.fr

**RECEIVED DATE**

Modeling the Effect of Wall Reactions on Oxidation

[\*]Corresponding author: DCPR-ENSIC, 1 rue Grandville, BP 20451, 54001 NANCY Cedex, France. E-mail: Richard.Porter@ensic.inpl-nancy.fr ; Tel.: 33 3 83 17 50 43 , Fax : 33 3 83 37 81 20

[†]E-mail: Frederique.Battin-Leclerc@ensic.inpl-nancy.fr

**ABSTRACT**

Simulations have been performed with spatial uniformity assumed, in order to assess the effect of tentative wall reactions on autoignition delay times and on a pressure - ambient temperature diagram of oxidation phenomena in the case of n-butane. Reactions which depend on the type of reactor wall coating have been considered for $HO_2$ and $RO_2$ radicals with estimated rate constants. The inhibiting effect of these walls has been simulated for closed vessel autoignition delay times as well as for ignition boundaries at pressures below 1 atm. The minimum temperature at which reaction occurs in jet-stirred reactor simulations is similarly affected.

## 1. INTRODUCTION

Previous review studies[1-2] have shown that the coating of the wall of the reactor can significantly influence the position of the autoignition boundary of hydrogen/oxygen and the boundaries between the different oxidation phenomena (slow reaction, cool flame, single or multiple stage autoignition) of organic compounds in a pressure - ambient temperature ($p$-$T_a$) ignition diagram. The effects of surface treatment on the internal wall of a 254 cm$^3$ cylindrical reaction vessel have been studied by Cherneskey and Bardwell.[3] The $p$-$T_a$ ignition diagrams for equimolar n-butane/oxygen in an untreated silica vessel, a vessel internally coated with NiO and a vessel internally coated with PbO, as shown in figure 1, reveal that ignition and cool flame boundaries are modified by the introduction of such coatings, with the greater inhibitory effects being observed with PbO. A marked increase in the pressures and temperatures required to bring about autoignition are observed for the NiO coated vessel with further increases for the vessel coated with PbO. The kinetic reason for the suppression of reactivity is believed to be the enhanced ability of the coatings to destroy free radical intermediates at the vessel wall. In the same study, vessels that had been thoroughly cleaned with nitric and sulphuric acid were also found to have inhibitory features relative to 'aged' vessels.

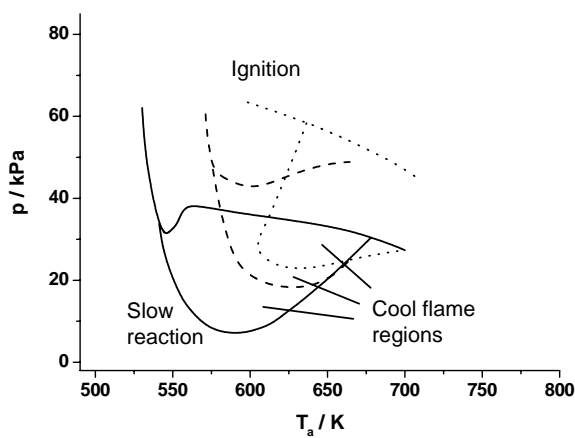

**Figure 1.** Experimental $p$-$T_a$ ignition diagrams for equimolar n-butane + oxygen mixtures (50 % n-butane and 50 % oxygen by volume without dilution) in a 254 cm$^3$ cylindrical vessel with different vessel internal wall coatings. Solid line - untreated silica vessel. Dashed line - NiO coated. Dotted line - PbO coated.[3]

Further studies into the effect of heterogeneous reactions on partial oxidation processes of methane have been conducted by Konnov et al..[4] Experimental results from an uncoated tubular glass reactor operated with long residence times were compared to a modeling study, in order to assess the impact of heterogeneous reactions, such as the adsorption of small radicals on the surface of the reactor tube wall. Mole fractions of products were quantitatively analysed over a range of temperatures. Simulations were carried out using selected values for sticking probabilities of the following reactions; adsorption of H atoms on the surface, adsorption of O atoms on the surface, reaction of $CH_3$ radicals with H atoms adsorbed on the surface, reaction of O atoms with O atoms adsorbed on the surface, and reaction of $CH_3O_2$ radicals with H atoms adsorbed on the surface. The gas-phase products of the last three reactions are methane, oxygen, and methyl peroxide ($CH_3O_2H$), respectively. Simulation results were found to give good agreement with the experiment at a number of operating temperatures. The authors concluded that some temperature dependence of the heterogeneous reactions was illustrated in the comparison of the experiment and modeling.

The purpose of the work presented in this paper is to define potentially important heterogeneous reactions with associated rate parameters and to qualitatively assess their effect on the predicted behaviour of a detailed gas phase model of the oxidation of n-butane, comprising 128 species in 1148 irreversible reactions. The effect of incorporating a number of different heterogeneous reactions is investigated via the comparison of closed vessel calculations of ignition delay times and sub-atmospheric $p$-$T_a$ ignition diagrams, as well as molecular species profiles in a jet-stirred reactor. Details about the origin of the comprehensive mechanism and its validation can be found in a recent paper.[5] Discrepancies between zero dimensional simulations using such mechanisms and experimental data from unstirred closed vessels at corresponding conditions may often be observed.[6] There are a number of possible reasons: Firstly, as the exothermic reaction begins near the centre of the vessel, buoyancy and convection will inevitably arise under normal gravity[7-10] and spatial inhomogeneities of temperature and concentration will exist. These effects cannot be incorporated into a spatially uniform model and pose significant challenges to more detailed physical modeling approaches.[9-10] The important effect of these forces on surface termination is not readily quantifiable, but convection is known to be the dominant mode of transport in the low temperature gas phase combustion of hydrocarbons.[9] Diffusion and convection rates in the experimental reactor under consideration are investigated in section 3.4. Secondly, there may be missing or incorrect reaction steps in the kinetic mechanisms, e.g., the lack of any heterogeneous wall reactions from many comprehensive mechanisms can have an effect on their exhibited reactivity. Furthermore, uncertainty in the kinetic rate or thermokinetic parameters can often lead to discrepancies. Large proportions of the parameters included in today's comprehensive mechanisms have never been studied experimentally or by ab initio modeling techniques and are often deduced from similar reactions or by thermochemical kinetic methods. The inaccuracy of these techniques can have a large bearing on simulated output. However, a recent global uncertainty analysis[11] of the effect of the uncertainties on ignition delay times and the positions of the boundaries in a $p$-$T_a$ ignition diagram at low temperature has shown that only a limited set of parameters may be responsible for most of the output variance, with the thermochemistry of $RO_2$ radicals having a large influence. In the context of cool flame experiments performed under microgravity conditions, zero-dimensional simulations with surface destruction under diffusion control of all radicals and peroxides have also been performed.[11] With the wall losses under pure diffusion control, reaction rates were typically 0.2 $s^{-1}$ for most species.

## 2. DEFINITION OF THE WALL REACTIONS AND THEIR RATE PARAMETERS

In this section, we briefly discuss the factors which influence the selection of important gas-phase species which require wall reaction modeling. Wall reactions are then defined with rate constants which depend on reactor geometry and surface activity but are temperature independent and not diffusion limited. The rate of wall consumption of a particular species will depend both on its mass transport and reactivity within the gas phase. Uncertainty analysis performed under low temperature autoignition conditions has shown that there is an insensitivity to the wall loss of small reactive radicals such as OH,

O and H, but that the surface termination of peroxide intermediates is much more influential, probably due to their slower reactivity relative to those of the most reactive species, such that diffusion to the surface becomes important.[11] According to previous work[12-15] amongst the radicals present in oxidation, only $HO_2$ radicals would be sufficiently unreactive to diffuse to the surface and promote heterogeneous reactions. However, at low temperature, $RO_2$ radicals will be in sufficient quantities for their wall reactions to also become potentially important. The diffusion coefficients for the species of interest have been derived in nitrogen, and for the case in the absence of diluent, in n-butane in the form

$$D_{AB} = \frac{0.00266 T^{3/2}}{p m_{AB}^{1/2} \sigma_{AB}^2 \Omega_D},$$

where $D_{AB}$ is the diffusion coefficient of species $A$ relative to species $B$, $T$ is temperature, $p$ is pressure, $m$ is reduced mass, $\sigma_{AB}$ is the Lennard-Jones collision diameter and $\Omega_D$ is the diffusion collision integral.[16] The Lennard-Jones parameters were not available for $RO_2$ radicals so the values of n-butane were assigned in their place on the basis of a similarly high molecular mass. The diffusion coefficients of the species pertaining to the initial conditions of 565 K, 40 kPa and equimolar n-butane and oxygen are given in table 1, along with their concentrations at $\Delta T = 1$ K. Although the diffusivity of $HO_2$ and $RO_2$ radicals is less than that of the smaller species (H atoms and OH radicals), this is offset by their more significant concentration which gives their wall reactions greater importance. The experimentally determined rate constants of Zils et al.[17] for the wall reactions of $HO_2$ radicals and H atoms show similar orders of magnitude in their reactor operated between 2.6 – 26 kPa and 773 – 793 K, so mass transfer and concentration in the gas phase are the greatest factors controlling the rate of wall loss of these species.

**Table 1.** Simulated concentration and diffusion coefficients of selected species. Concentrations pertain to equimolar n-butane and oxygen at $T_a = 565$ K, $p = 40$ kPa and $\Delta T = 1$ K. Diffusivities correspond to the initial conditions.

|  | OH | H | HO$_2$ | RO$_2$ |
|---|---|---|---|---|
| Concentration / molecules cm$^{-3}$ | 5.4×10$^8$ | 7.4×10$^8$ | 2.7×10$^{12}$ | 5.7×10$^{13}$ |
| $D_{\text{(n-butane)}}$ / cm$^2$ s$^{-1}$ | 1.29 | 5.23 | 0.82 | 0.35 |
| $D_{\text{(nitrogen)}}$ / cm$^2$ s$^{-1}$ | 2.38 | 9.34 | 1.57 | 0.79 |

Cheaney et al.[14] studied the effects of surfaces on combustion of methane and divided surfaces into two classes, each related to a specific heterogeneous reaction:

- Category I includes surfaces treated with acid. The fate of HO$_2$ radical is assumed to be reduction to H$_2$O$_2$ by analogy with acid solutions and is related to reaction (W1):

$$\text{HO}_2 \text{ (g)} + \text{H}^+ \text{ (s)} \rightarrow \tfrac{1}{2} \text{H}_2\text{O}_2 \text{ (g)} + \tfrac{1}{2} \text{O}_2 \text{ (g)} + \text{H}^+ \text{ (s)} \qquad (\text{W1})$$

Which may be written in two steps for modeling purposes as:

$$\text{Vacancy (s)} + \text{HO}_2 \text{ (g)} \rightarrow \text{HO}_2 \text{ (s)} \qquad (\text{W1-a}),$$

$$2 \text{ HO}_2 \text{ (s)} \rightarrow \text{H}_2\text{O}_2 \text{ (g)} + \text{O}_2 \text{ (g)} + 2 \text{ Vacancy (s)} \qquad (\text{W1-b}).$$

- Category II includes surfaces coated with salt or metal oxides. The failure to detect H$_2$O$_2$ in reaction carried out with category II surfaces led Cheaney et al.[14] to propose another fate for HO$_2$ radicals leading to more stable species and is related to reaction (W2).

$$\text{HO}_2 \text{ (g)} + \text{e}^- \text{ (s)} \rightarrow \tfrac{1}{2} \text{H}_2\text{O} \text{ (g)} + \tfrac{3}{4} \text{O}_2 \text{ (g)} + \text{e}^- \text{ (s)} \qquad (\text{W2}).$$

Alternatively as:

$$\text{Vacancy (s)} + \text{HO}_2 \text{ (g)} \rightarrow \text{HO}_2 \text{ (s)} \qquad (\text{W2-a}),$$

$$4 \text{ HO}_2 \text{ (s)} \rightarrow 2 \text{ H}_2\text{O} \text{ (g)} + 3 \text{ O}_2 \text{ (g)} + 4 \text{ Vacancy (s)} \qquad (\text{W2-b}).$$

In these and other cases, a first order dependence was achieved by arbitrarily setting a high rate constant for step (b), and so step (a) becomes rate limiting as a result. As predicted by these reactions, H$_2$O$_2$ was experimentally observed when the surfaces were of category I, but not when they were of category II. Since this implies the destruction of all H$_2$O$_2$ in category II vessels, and because its formation from the reaction between HO$_2$ radicals and hydrocarbon molecules cannot be neglected, an additional reaction has been proposed for surfaces of category II by Cheaney et al.[14] by analogy with catalytic mechanisms proposed on metal surfaces:

$$\text{H}_2\text{O}_2 \text{ (g)} + \text{e}^- \text{ (s)} \rightarrow \text{H}_2\text{O} \text{ (g)} + \tfrac{1}{2} \text{O}_2 \text{ (g)} + \text{e}^- \text{ (s)} \qquad (\text{W3}),$$

or,

$$\text{Vacancy (s)} + H_2O_2 \text{ (g)} \rightarrow H_2O_2 \text{ (s)} \qquad\qquad \text{(W3-a)},$$

$$2\, H_2O_2 \text{ (s)} \rightarrow 2\, H_2O \text{ (g)} + O_2 \text{ (s)} + 2\, \text{Vacancy (s)} \qquad\qquad \text{(W3-b)}.$$

These three reactions are taken into account in our modeling study, with rate parameters estimated as explained below and with the assumption of rate control by surface activity. That is, the surface loss of radicals is not under diffusion control and rates are determined from bulk gas concentrations in the zero-dimensional model. Expressions of rate constants for the removal of radicals at the wall in chain reactions have been given by Blackmore.[18] At high Knudsen numbers or when the profile of radicals in the reactor is flat, as may occur if the diffusion rate is high and/or the wall destruction rate is weak,[19] these expressions are given as;

$$k_w = \frac{\gamma \bar{c}}{d} \quad \text{for a cylindrical reactor}$$

and

$$k_w = \frac{3}{2}\frac{\gamma \bar{c}}{d} \quad \text{for a spherical reactor.}$$

where $\gamma$ is the probability of reaction on each collision with the surface, $\bar{c}$ is the radicals average molecular velocity which can be obtained from the kinetic theory of gases (at an average temperature of 800 K, for $\bar{c} \approx 71700$ cm.s$^{-1}$ for $HO_2$ radicals and $H_2O_2$ molecules) and d is the diameter of the reactor. Strictly, these expressions have been derived in the absence of forced or natural convection. However, since they are applied to reactors with uniform concentration they can be extrapolated to well mixed scenarios, but may give conservatively large estimates in unstirred vessels where natural convection prevails. In the case where diffusion is limiting, $k_w$ would be proportional to the diffusivity $D$ which is a function of $m_{AB}^{-1/2}$ as $\bar{c}$ is, so the relative rate constants between $HO_2$ radicals, $H_2O_2$ and $RO_2$ radicals might be preserved.

As the influence of $HO_2$ radicals is less important than that of alkylperoxy radicals ($RO_2$) on the low temperature chemistry of the oxidation of hydrocarbons, it is important to test the effect of category II surface reactions involving these last radicals. In the case of RO$_2$ radicals, the following reaction can be derived from those proposed for HO$_2$ radicals with the same assumption for the rate constant estimation:

$$RO_2 \text{ (g)} + e^- \text{ (s)} \rightarrow \text{½ RH (g)} + \text{½ conjugated olefin (g)} + O_2 \text{ (g)} + e^- \text{ (s)} \quad\quad (W4),$$

or,

$$\text{Vacancy (s)} + RO_2 \text{ (g)} \rightarrow RO_2 \text{ (s)} \quad\quad (W4\text{-a})$$

$$2\ RO_2 \text{ (s)} \rightarrow RH \text{ (g)} + \text{conjugated olefin (g)} + 2\ O_2 \text{ (g)} + 2\text{Vacancy (s)} \quad\quad (W4\text{-b}).$$

Here, $\bar{c} \approx 43600$ cm.s$^{-1}$ for butyl peroxy radicals. Stable products have been assumed to be alkane and conjugated alkene rather than ether since they are already included in the mechanism. The nature of the molecular products is not sensitive because of the small flow rate of the termination step W4.

## 3. RESULTS AND DISCUSSION

**3.1. Ignition Delay Times.** Figures 2 and 3 present the influence of wall effects on simulated autoignition delay times for stoichiometric n-butane/air mixtures in a spherical reactor (10 cm diameter). Simulations were performed adiabatically using SENKIN[20] and the gas-phase mechanism of Buda et al.[5] for the following scenarios;

(i) in the absence of wall termination,

(ii) with wall reaction (W1),

(iii) with wall reactions (W2) and (W3),

(iv) with wall reaction (W4),

(v) with wall reactions (W2), (W3) and (W4).

For the reactions of HO$_2$ radicals, we have first assumed a low probability of reaction, $\gamma_{HO2}$, of 0.01 corresponding to a very light treatment of the surface. Zils et al.[12] derive from their experiments values of $\gamma_{HO2}$ between 0.017 and 0.062 according to the efficiency of treatment of the reactor with PbO. We have then obtained $k_{w1a} = k_{w2a}$ equal to 107 s$^{-1}$ and, $k_{w3a} = 0.5$ s$^{-1}$, assuming the same relationship for H$_2$O$_2$ molecules with a much lower probability of reaction, as they are not radicals, $\gamma_{H2O2} = 5\times10^{-5}$ (i.e.

200 times lower than $\gamma_{HO2}$). Zils et al.[12] report a low activation energy of 8 kcal mol$^{-1}$ for the reaction of HO$_2$ radicals with the surface, which was determined over a very narrow temperature range of 773-793 K, so a degree of uncertainty resides in this value. Because our investigation focuses on rather narrow ranges of temperature, it is appropriate to specify the rate parameters solely as A-factors in order to prevent any additional uncertainty, arising from the activation energy, affecting our model. Figure 2 shows that the effect of heterogeneous reactions depends strongly on pressure. At 5 bar, this effect is weak whatever the type of reaction, whereas at 0.5 bar, the delay times are multiplied by a factor of up to 10 when considering reaction (W1) and by a factor of more than 1000 when considering wall reactions (W2) and (W3). The quantitative distinction of the behaviour at higher pressure arises here solely because the concentration dependences of the bimolecular and termolecular reactions are favoured more strongly than the linear termination reaction. Since the inverse pressure dependence of diffusion controlled reactions is neglected, the influence of heterogeneous reactions is certainly overestimated at higher pressure in these calculations. Simulations of low temperature combustion reactions taking into account both a detailed chemical model and 1D heat and mass diffusion are still scarce.[21] In addition, under normal gravity, convection effect should also be considered. Since $k_{w1a} = k_{w2a}$, the surface destruction of HO$_2$ must be of equal importance in the predicted behaviour whatever the type of surface. Thus the difference seen in figure 2b below 750 K between wall reaction (W1) and wall reactions (W2 and W3) must be attributed to the supplementary effect of surface destruction of H$_2$O$_2$, as far as the value used for $\gamma_{H2O2}$ is somewhat meaningful. Furthermore, we have considered a higher probability of reaction, $\gamma_{HO2}$, of 0.1 for wall reaction (W1), corresponding to a much more strongly treated reactor. Figure 2 shows that with this higher value, wall reactions have a marked effect on the ignition delay times, even at 5 bar.

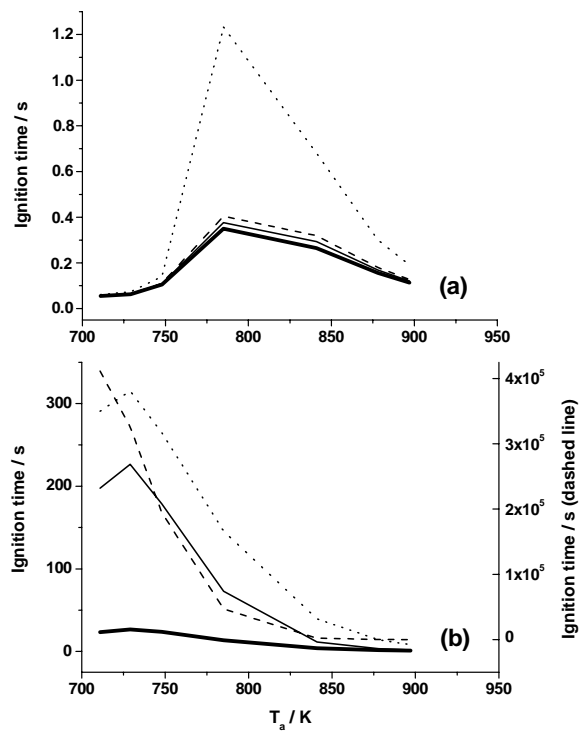

**Figure 2.** Influence of wall effects on simulated autoignition delay times for stoichiometric n-butane/air mixtures for an initial pressure of (a) 5 bar and (b) 0.5 bar. Thick lines correspond to the absence of wall termination, thin lines to wall reaction (W1), dashed lines to wall reactions (W2 and W3) and dotted lines to wall reaction (W1) with a higher probability of reaction ($\gamma = 0.1$).

Experimental data of n-butane autoignition with initial temperatures between 700-800 K was obtained from untreated glass vessels of varying size (100 ml, 200 ml and 500 ml) by Pekalski[22]. This may be used to verify the simulation results shown in figure 2 because as vessel size decreases, surface to volume ratio increases, thus providing a more accessible surface for termination. Qualitatively, the results agree well; the smallest experimental vessel shows the greatest reaction inhibition at 1 atm. Increasing the pressure to 10 bar in the experiment also demonstrated a significant shortening of ignition delay times.

For the reaction of $RO_2$ radicals (W4), we have considered the same probability of reaction as for $HO_2$ radicals: $\gamma_{RO2} = 0.01$, corresponding to $k_{w4a} = 65.4$ s$^{-1}$ ($\bar{c} \approx 43600$ cm.s$^{-1}$ for butyl peroxy radicals).

Figure 3 shows that the effect of reaction (W4) on ignition delay times is much more negligible than the others except at low pressure and temperatures below 750 K, where the rates of competing temperature dependent reactions are reduced and the concentration of $RO_2$ radicals is higher. The concentration of $RO_2$ radicals in a jet-stirred reactor is discussed further in the text.

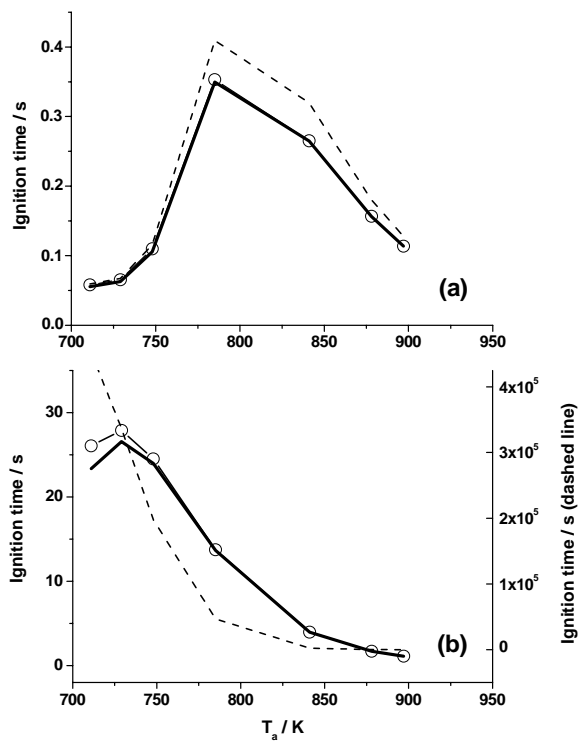

**Figure 3.** Influence of wall effects on simulated autoignition delay times for stoichiometric n-butane/air mixtures for an initial pressure of (a) 5 bar and (b) 0.5 bar. Thick lines correspond to the absence of wall termination, open circles to wall reaction (W4) and dashed lines to wall reactions (W2), (W3) and (W4).

**3.2. $p\text{-}T_a$ Ignition Diagrams.** Figure 4 displays simulated $p\text{-}T_a$ ignition diagrams of oxidation phenomena for equimolar n-butane/oxygen mixtures obtained under the conditions of Cherneskey and Bardwell[3] with and without the two types of wall reactions related to the consumption of $HO_2$ and $H_2O_2$. In this case, the reactor is cylindrical (6 cm diameter) and has been strongly treated involving a high probability of reaction: $\gamma_{HO2} = 0.06$ and $\gamma_{H2O2} = 3\times10^{-4}$, corresponding to $k_{w1a} = k_{w2a} = 717$ s$^{-1}$ and $k_{w3a} = 3.6$ s$^{-1}$. Heat loss through the walls is described via Newtonian cooling. An overall heat transfer coefficient of $3 \times 10^{-3}$ W cm$^{-3}$ K$^{-1}$ was found to be appropriate to the pyrex vessel during the simulation.

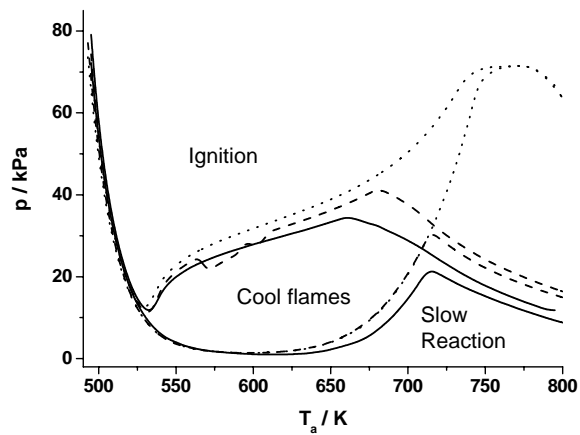

**Figure 4.** Influence of $HO_2$ related wall reactions on the $p$-$T_a$ ignition diagram of oxidation phenomena for equimolar n-butane/oxygen mixtures. Solid line – no wall effect. Dashed line – wall reaction (W1). Dotted line – wall reactions (W2) and (W3).

These diagrams were constructed automatically using a software developed at the University of Leeds[23] and based on UNIX shell scripts to control the execution of a modified version of SPRINT[24] in which the various non-isothermal behaviour i.e., ignition, cool flames, and slow reaction are characterised. This characterisation of the reaction modes is based on monitoring the temperature increase and temperature gradient within the simulated time. Appropriate criteria were defined for the characterisation such that during slow reaction $\Delta T$ does not exceed 20 K. For low initial temperatures, ignition is deemed to have occurred if $T$ exceeds 1100 K. The identification of a cool flame or multiple cool flames requires not only that $\Delta T > 20$ K, but also that a maximum is identified before T exceeds 1100 K. Individual simulations were terminated after ignition had occurred or after 4000 s reaction time. If no cool flames had been detected before this time then the behaviour would be classified as a slow reaction.

The diagram obtained without considering heterogeneous reactions is qualitatively similar to that observed experimentally in figure 1, even if quantitative differences are encountered for the position of some boundaries, e.g. the simulated minimum ignition temperature is about 515 K and the simulated pressure ignition boundary between 550 - 650 K ranges between 10 and 30 kPa, whereas the experimental minimum ignition temperature is about 530 K and the experimental pressure ignition boundary between 550 - 650 K varies between 30 and 40 kPa.

Simulations considering heterogeneous reactions related to $HO_2$ do not show changes in the minimum ignition temperature as encountered by Cherneskey and Bardwell,[3] but display well an increase of the pressure ignition boundary above 550 K. Below 550 K, the influence of $HO_2$ radicals is less important than that of alkylperoxy radicals ($RO_2$) and wall reactions of these last radicals should probably be taken into account to reproduce a change in the minimum ignition temperature. For wall reactions (W2-W3), the pressure ignition boundary between 550 K and 670 K increases by only 2.9 to 8.9 kPa, with a slightly less pronounced effect for wall reaction of (W1). However, further increases at higher temperatures are noticed, with the inhibiting effect of the wall reactions (W2-W3) above 670 K being much greater. Qualitatively, this trend is observed in the experimental results[3] to a limited extent for reactors coated with NiO.

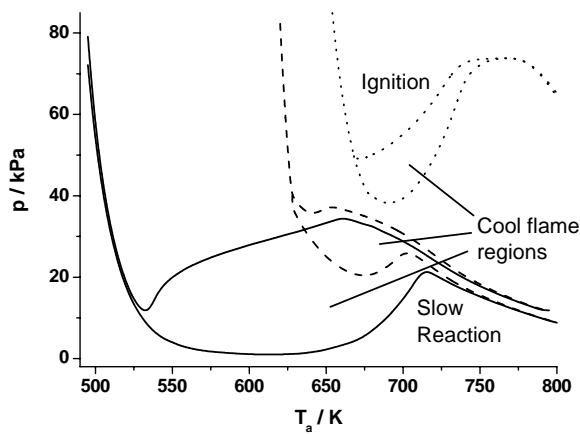

**Figure 5.** Influence of $RO_2$ and $HO_2$ related wall reactions on the $p$-$T_a$ ignition diagram of oxidation phenomena for equimolar n-butane/oxygen mixtures. Solid line – no wall effect. Dashed line – wall reaction (W4). Dotted line – wall reactions (W2), (W3) and (W4).

Figure 5 displays simulated $p$-$T_a$ ignition diagrams of oxidation phenomena for equimolar n-butane/oxygen mixtures obtained under the conditions of Cherneskey and Bardwell[3] with wall reaction (W4) for $RO_2$ radicals and for the combined effect of wall reactions (W2, W3 and W4) relating to $RO_2$ and $HO_2$ radicals, and $H_2O_2$. In the cylindrical reactor, we consider the same probability of reaction as for $HO_2$ radicals: $\gamma_{RO2} = 0.06$, corresponding to $k_{w4a} = 436$ s$^{-1}$ ($\bar{c} \approx 43600$ cm.s$^{-1}$ for butyl peroxy radicals). The position of the low temperature ignition boundary is kinetically controlled to a large extent by the $R + O_2 / RO_2$ equilibrium. The slow reaction region is controlled by an overall non-branching reaction mode, with the termination rate exceeding the chain branching rate. In the ignition region, the equilibrium is displaced towards the formation of $RO_2$ and degenerate branching is predominant.[25] An increased consumption of $RO_2$ radicals by the reactor wall will decrease the reactivity in this region of the *p-T$_a$* diagram, shifting the ignition boundary to a position of higher temperature. This kinetic phenomenon is qualitatively reflected in the modeling by figure 5, with the low temperature ignition boundary being shifted to around 625 K for wall reactions of $RO_2$. A less pronounced pressure increase of the ignition boundary is observed above 650 K. Nevertheless, the minimum pressure at which cool flames occur, at 675 K is increased by some 15 kPa. The *p-T$_a$* diagram generated using the mechanism incorporating wall reactions (W2, W3 and W4) related to both $HO_2$ and $RO_2$ radicals displays many of the qualitative characteristics observed experimentally, with shifts of all boundaries to positions of higher temperature and pressure. This substantiates the significance of the effect of wall reactions, related to both of these radicals, on the positions of the boundaries in the *p-T$_a$* ignition diagram. Below 700 K, the main effect is caused by $RO_2$ radicals, while above 700 K it can be largely attributed to $HO_2$ radicals. A rate of production analysis conducted with this mechanism at conditions corresponding to 670 K and 60 kPa on the ignition diagram shows that wall loss reactions account for 2 – 10 % of the destruction rate of $RO_2$ radicals during the ignition delay. A greater proportion of $HO_2$ radicals are consumed by the wall with a 93 % destruction rate for this reaction shortly after initiation (t = 0.003 s). This value falls dramatically to 27 % prior to two-stage ignition (t = 1.55 s) when the temperature has climbed to 680 K.

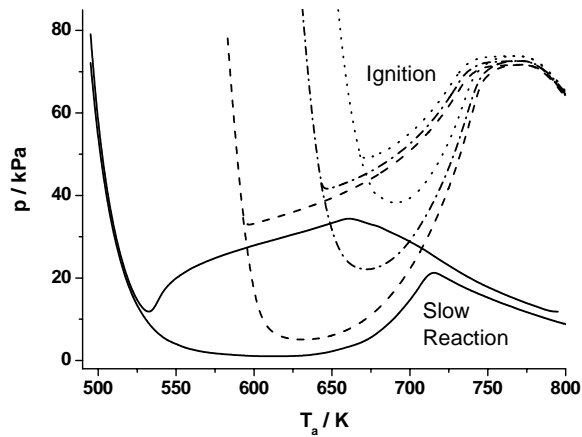

**Figure 6.** Influence of the magnitude of the rate parameter of $RO_2$ related wall reactions on the $p$-$T_a$ ignition diagram of oxidation phenomena for equimolar n-butane/oxygen mixtures. Solid line – no wall effect. Dotted line – wall reactions (W2, W3 and W4), $k_{w4a} = 436$ s$^{-1}$. Dashed line – wall reactions (W2, W3 and W4), $k_{w4a} = 43.6$ s$^{-1}$. Dotted-dashed line – wall reactions (W2, W3 and W4), $k_{w4a} = 218$ s$^{-1}$.

Further $p$-$T_a$ ignition diagrams have been constructed using the mechanism incorporating reactions (W2), (W3) and (W4) but with adjusted rate coefficients for the latter $RO_2$ wall reaction in order to assess the influence of its magnitude on the position of the low temperature ignition boundary. The introduction of the wall reactions with unadjusted rate coefficients raises the ignition boundary from a temperature of 500 K to a position at 660 K. Figure 6 shows the result of dividing the rate constants for the wall reaction (W4) by a factor of 2 or 10. Dividing by 2 results in a decrease in the position of the autoignition boundary by 25 K, whereas a division by 10 decreases the value more strongly to 590 K, bringing it close to the experimental value[3] for an NiO wall coated vessel, shown in figure 1.

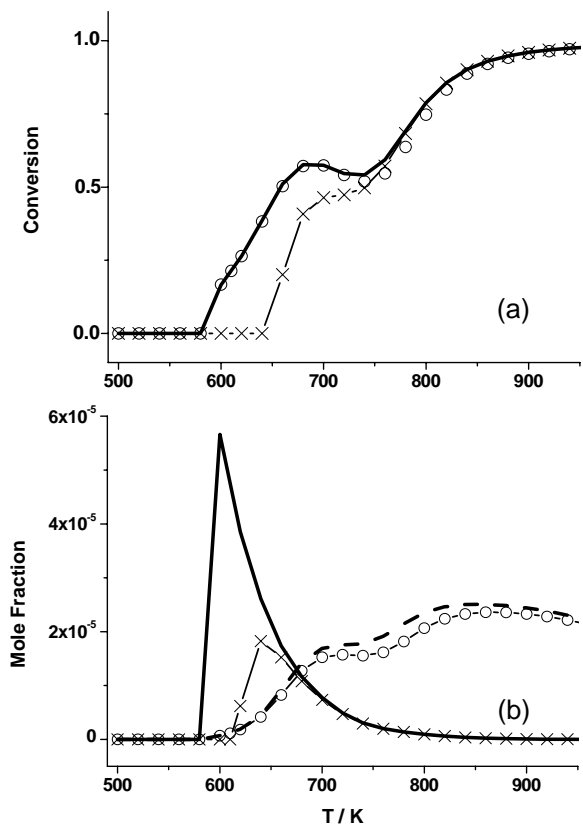

**Figure 7.** Influence of wall reactions on the molecular species profiles obtained during the numerical simulation of the oxidation of a stoichiometric n-butane/air mixture, at 10 atm pressure and 0.5 s mean residence time. Figure (a); **conversion profiles**: Thick line – no wall effect. Open circles – wall reactions (W2 and W3). Crosses – wall reaction (W4). Figure (b); **HO$_2$ radical profiles**: Dashed line – no wall effect. Open circles – wall reactions (W2 and W3). **RO$_2$ radical profiles**: Thick line – no wall effect. Crosses – wall reaction (W4). In the last two cases, mole fractions of RO$_2$ isomers have been summed.

**3.3. Jet-Stirred Reactor.** Experimental jet-stirred reactors[26] are designed according to strict criteria so that mass and thermal gradients are eliminated. The kinetic energy of the inlet gas entering the reactor through three or four nozzles enables good mixing in the gas phase except for in a thin boundary layer close to the vessel wall. Figure 7 shows the results of a numerical investigation of oxidation in a jet-stirred reactor using the original comprehensive mechanism, one incorporating reactions (W2) and (W3) relating to HO$_2$ radicals and one incorporating wall reaction (W4) of RO$_2$ radicals. A spherical jet-stirred reactor may typically have a diameter of 4 cm and for this study we have assumed the same probabilities of reaction used in the autoignition delay times study in a spherical reactor: $\gamma_{HO2} = \gamma_{RO2} = 0.01$ and $\gamma_{H2O2} = 5 \times 10^{-5}$. These correspond to the rate constant values of $k_{w2a} = 269 \text{ s}^{-1}$, $k_{w3a} = 1.34 \text{ s}^{-1}$ and $k_{w4a} = 65.4 \text{ s}^{-1}$. Simulations were carried out using the PSR module of CHEMKIN[20] under stoichiometric conditions in air, over the low temperature range, at 10 atm and using a mean residence time of 0.5 s. The simulated mole fraction profiles of figure 7 show that the introduction of $RO_2$ wall reactions results in an increase in the minimum temperature at which oxidation occurs, a slight decrease in $HO_2$ radical mole fraction and a large decrease in the mole fraction of $RO_2$ radicals at temperatures below 700 K where they are more abundant. The reactivity is therefore decreased at these operating conditions. The $HO_2$ related reactions have a much more limited effect on the profiles of their radicals and fuel – most noticeable between 700 – 850 K. These results agree well with the others reported in this paper.

**3.4. Order of Magnitude Calculation for Diffusion and Convection in Experimental Reactors**

The purpose of this section is to investigate heat and mass transport properties in the experimental closed vessel reactor[3] under consideration in this study which is cylindrical, 6 cm in diameter, 9 cm long and held in the vertical position.[27] The simulations reported in this paper were conducted with zero-dimensional codes, which are commonly employed for investigations of alkane autoignition but assume spatial uniformity. As non-isothermal reaction evolves in a real reactor, there may be spatial variations of temperature and concentration.[11] Cherneskey and Bardwell give no details regarding the convection or spatial variations of temperature in their reactor, but other experimentalists report that self heating occurs near the centre, with pre-cool flame temperature rises of 10 – 20 K,[28] resulting in convection during the induction period. The buoyant hot gas rises in the bulk and recirculates downward along the internal vessel wall as it cools. The cool flame develops in a hot buoyant plume above the centre of the vessel.[10,27] Convection and diffusion rates arising at the conditions of interest might provide an adequate degree of transport to the reactor walls so that the assumption of spatial uniformity is reasonable. The degree of convection is critical and conditions under which turbulent convection prevails correspond directly to the assumption of spatial uniformity. The length of reaction time ($\tau_{reaction}$) will also have a large bearing on the degree of uniformity. We aim to test these effects via the calculation of convection and diffusion timescales using the initial conditions of $T_a$ = 565 K and $p$ = 40 kPa as the test case. To do this, we use a number of dimensionless numbers and associated formulae for mass and heat transport in the experimental reactor (see Appendix).[29]

The characteristic length used for the diffusion calculations is the diameter of a sphere of equivalent $V / S$ ratio. A value for the Fourier number for mass transport $Fo_m$ = 1 and a diffusion coefficient for $HO_2$ = 0.82 cm$^2$ s$^{-1}$ yields a characteristic diffusion time ($\tau_{diffusion-m}$) of 55 s. Similarly for heat diffusion, we get a characteristic diffusion time ($\tau_{diffusion-h}$) of 74 s. Comparison of these values to the experimental ignition delay time in the untreated vessel, here ~ 18 s, implies that heat and mass diffusion alone is unlikely to fully smooth out spatial inhomogeneities. Turning to convection, using a characteristic length in the direction of gravity (9 cm) and assuming a conservative pre-cool flame $\Delta T$ of 1 K between the bulk gas and the reactor wall, we can estimate the Reynolds number $Re$ via the Grashof number $Gr$. This leads to $Re$ = 39, a gas velocity of 0.03 m s$^{-1}$ and a characteristic convection time ($\tau_{convection}$) of 3.7 s. Figure 8 shows that $\tau_{convection}$ varies little when $\Delta T$ exceeds 5 K. Very similar curves are obtained at other initial conditions near the ignition boundaries. $Re$ scales with $\Delta T$ but does not exceed its critical value during the induction period. Turbulent convection may occur in larger vessels and at higher pressures.[7-9] At differing experimental conditions ($T_a$ = 557 K and $p$ = 32 kPa) two consecutive cool flames occur with an induction time of the first one of 90 s.[3] In this case $\tau_{reaction}/\tau_{convection}$ ~ 24.3, so a greater extent of mixing will have been achieved compared to the ignition case. With such order of magnitude values for the diffusion and convection timescales it is difficult to say categorically that the assumption of a uniform gas mixture in the untreated vessel is perfectly valid well before the onset of the cool flame, but it is likely that a good degree of mixing has taken place and this will increase as the temperature further rises.

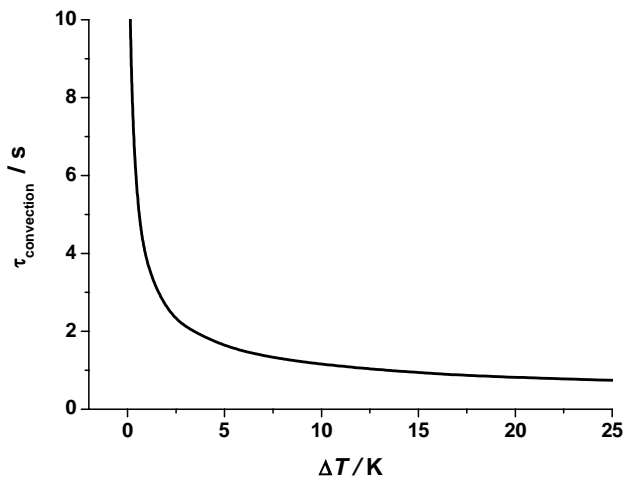

**Figure 8.** Estimate of the convection timescale as a function of the temperature excess at the centre of the reactor for a vertical cylinder 9 cm long and 6 cm in diameter containing equimolar n-butane + oxygen at $T_a$ 565 K and $p = 40$ kPa.

As we have seen, the introduction of metal oxide coatings to the reactor walls can have a marked effect on the induction times. From a modeling perspective, transport to the wall for chemical reaction is more important under treated conditions. At $T_a = 565$ K and $p = 40$ kPa, the introduction of a KI coating to the experimental reactor wall increases $\tau_{\text{reaction}}$ from 18 s to 120 s (KI generally has a slightly weaker inhibiting effect than NiO). Moreover, when the wall is treated with PbO, $\tau_{\text{reaction}}$ increases to over 1600 seconds (slow reaction). Given such induction times in the treated cases, there seems little reason to doubt that the assumption of a well mixed vessel is reasonable because there is ample time for mixing by convection to occur. However, as the gas mixtures do not become turbulent in the unstirred closed vessels, the predictions obtained of wall effects on the $p$-$T_a$ diagram using the rate constants selected here may be overestimated somewhat in this limiting case assumption of spatial homogeneity.

## 4. CONCLUSIONS

Using previous studies as a basis, important heterogeneous reactions, relating to reactors whose internal walls have been treated with acid or metal oxides, have been identified and associated rate parameters estimated. Simulations have been performed using a comprehensive mechanism for the gas phase oxidation of n-butane and a range of mechanisms which incorporate the heterogeneous reactions in order to assess their effect on predicted autoignition behaviour. Results show that wall reactions relating to $HO_2$ radicals have a significant effect on simulated ignition delay times at sub-atmospheric pressure. At higher pressures their effect is less pronounced. Wall reactions relating to $RO_2$ radicals only have a significant effect on ignition delay times at low pressure and temperature below 700 K. Further simulations have been conducted using software to automatically construct the $p$-$T_a$ ignition diagrams of the mechanisms. Comparison to the experimentally determined $p$-$T_a$ diagram under the conditions where no treatment has been introduced to the reactor surface shows that a good qualitative prediction is displayed by the comprehensive mechanism, although quantitative discrepancies exist which may be a result of spatial inhomogeneities in the unstirred vessel or uncertainties in the mechanism. The $p$-$T_a$ ignition diagrams generated by mechanisms incorporating heterogeneous reactions related to $HO_2$ radicals show that these reactions increase the minimum pressure at which autoignition occurs but not the minimum temperature, and so are able to reproduce some but not all of the qualitative trends observed in the experiments. However, introducing heterogeneous reactions related to $RO_2$ radicals into the mechanism does have a significant effect on the minimum autoignition temperature and a combination of both $HO_2$ and $RO_2$ related wall reactions gives a good qualitative reproduction of the experimentally determined effect of metal oxide internal wall coatings on $p$-$T_a$ ignition diagrams. Additional studies have addressed the significance of the magnitude of the value of the rate parameters of $RO_2$ related wall reactions and have demonstrated similar inhibitory effects of these reactions in a jet-stirred reactor.

**Acknowledgement.** Financial support of this work by the European Union within the "SAFEKINEX" Project EVG1-CT-2002-00072 and of the PNIR "Carburants et moteurs" of CNRS. John Griffiths, Alison Tomlin and Gabriel Wild are thanked for helpful discussions. Cecile Mazelin is thanked for his contribution to a preliminary bibliographic study.

**APPENDIX**

(a) Formulae of Fourier Numbers

for chemical species diffusion

$$Fo_m = \frac{D\tau_{diffusion-m}}{L^2}$$

for thermal diffusion

$$Fo_h = \frac{\alpha\tau_{diffusion-h}}{L^2}$$

where

$D$ = diffusion coefficient (m$^2$ s$^{-1}$)

$\alpha$ = thermal diffusivity (m$^2$ s$^{-1}$)

$\tau_{diffusion}$ = characteristic diffusion time (s)

$L$ = Characteristic length (m)

(b) Calculation of the Grashof number

$$Gr = \frac{g\beta L^3 \rho^2}{\mu^2}\Delta T$$

where

$g$ = acceleration due to gravity (m s$^{-2}$)

$\beta$ = coefficient of cubical expansion (K$^{-1}$)

$\rho$ = density (kg m$^{-3}$)

$\mu$ = dynamic viscosity (kg m$^{-1}$ s$^{-1}$)

$\Delta T$ = temperature difference between the bulk gas and the vessel wall (K)

(c) Estimation of the Reynolds Number[28]

$$Re \cong \sqrt{\frac{Gr}{2.5}}$$

(d) Calculation of the gas velocity $u_g$ (m s$^{-1}$)

$$u_g = \frac{Re\,\mu}{L\rho}$$

(e) Calculation of the convection timescale $\tau_{convection}$ (s$^{-1}$)

$$\tau_{convection} = \frac{L}{u_g}$$

(f) Evaluation of $\beta$ for an ideal gas[7]

$$\beta = \frac{1}{T}$$

(g) Evaluation of $\mu$ and $\alpha$ at equimolar proportions of n-butane and oxygen, 550 K and 60 kPa[16]

$\mu = 1.79 \times 10^{-5}$ kg m$^{-1}$ s$^{-1}$

$\alpha = 3.95 \times 10^{-5}$ m$^2$ s$^{-1}$